%Paper: hep-ph/9312267
%From: yoshida@gauge.scphys.kyoto-u.ac.jp
%Date: Tue, 14 Dec 93 00:45:15 +0900
%Date (revised): Tue, 14 Dec 93 12:19:37 +0900
%Date (revised): Wed, 15 Dec 93 20:04:13 +0900

%%%%%%%%%%%   READ ME  READ ME  READ ME  READ ME  READ ME   %%%%%
%                                                               %
%    This paper has seven figures appended in a second part     %
%    as a uuencoded compressed tar file with instructions for   %
%    unpacking (Please search for '#!').                        %
%                                                               %
%%%%%%%%%%%%%%%%%%%%%%%%%%%%%%%%%%%%%%%%%%%%%%%%%%%%%%%%%%%%%%%%%
\input phyzzx
\input epsf
\input tables

\font\twelvemib=cmmib10 scaled\magstep1     \skewchar\twelvemib='177
\def\mib#1{\hbox{\twelvemib #1}}

\def\bfgamma{\gamma\kern-2.27mm\gamma\kern-2.27mm\gamma\kern-2.27mm\gamma}
\def\bfepsilon{\epsilon\kern-1.85mm\epsilon\kern-1.85mm\epsilon}
\def\bftheta{\theta\kern-1.97mm\theta\kern-1.97mm\theta}
\def\bfsigma{\sigma\kern-2.37mm\sigma\kern-2.37mm\sigma}

%\mathchardef\bfgamma  ="090D
%\mathchardef\bfepsilon="090F
%\mathchardef\bftheta  ="0912
%\mathchardef\bfsigma  ="091B
%\mathchardef\bfgamma  ="0B0D
%\mathchardef\bfepsilon="0B0F
%\mathchardef\bftheta  ="0B12
%\mathchardef\bfsigma  ="0B1B
%\def\bfgamma{{\vec \gamma}}
%\def\bfepsilon{{\vec \epsilon}}
%\def\bftheta{{\vec \theta}}
%\def\bfsigma{{\vec \sigma}}
%%%%%%%%%%%%%%%%%%%%%  PARTICLE.TEX  %%%%%%%%%%%%%%%%%%%%%%
\catcode`@=11

\newtoks\myphone      \myphone={3823}
\newtoks\bitnet       \bitnet={JPNYITP}
\def\KYOTOHEAD{\setbox0=\vbox{\baselineskip=10pt
         \ialign{\ninerm ##\hfil&\ninerm ##\hfil\cr
         \hskip -17pt Bitnet&~~\the\bitnet\cr
         \hskip -17pt Telephone&~~{\sixrm +}81-75-753-\the\myphone\cr
         \hskip -17pt Fax&~~{\sixrm +}81-75-753-3886\cr
         \hskip -17pt Telex&~~5422829 KUNSDP J\cr}}
               {\titlestyle{\fourteenrm DEPARTMENT OF PHYSICS\break
         \twelverm KYOTO UNIVERSITY\break
         \tenrm KYOTO 606, JAPAN\break
         \baselineskip=12pt}}
   \line{\hskip -7pt\box2\hfil\box0}}
\newtoks\KUNS
\newtoks\HETH
\newtoks\monthyear
\Pubnum={KUNS~\the\KUNS\cr HE(TH)~\the\HETH}
\monthyear={\monthname,\ \number\year}
\def\p@bblock{\begingroup \tabskip=\hsize minus \hsize
   \baselineskip=1.5\ht\strutbox \topspace-2\baselineskip
   \halign to\hsize{\strut ##\hfil\tabskip=0pt\crcr
   \the\Pubnum\cr \the\monthyear\cr }\endgroup}
\def\title#1{\vskip\frontpageskip \centerline{\fourteenpoint\bf #1}
		\vskip\headskip }
\def\Kyoto{\address{Department of Physics,~Kyoto University \break
                            Kyoto~606,~JAPAN}}

\def\andjournal#1&#2(#3){\begingroup \let\journal=\dummyj@urnal
    \sl #1\unskip~\bf\ignorespaces #2\rm
    (\afterassignment\j@ur \count255=#3) \endgroup\ignorespaces }
\def\figmark#1{\par\vskip0.5cm{\hbox{\centerline{
           \vbox{\hrule height0.5pt
           \hbox{\vrule width0.5pt\hskip4pt
           \vbox{\vskip5pt\hbox{Fig.#1}\vskip4pt}
           \hskip4pt\vrule width0.5pt}
           \hrule height0.5pt}}}}\par\noindent\hskip0.1cm\hskip-0.1cm}
\def\ee{\eqno\eq }
\def\abs#1{{\left\vert #1 \right\vert}}
%
%\nopubblock    %This command deletes the preprint number block.
hep-ph/9312267
\KUNS={1234}     %KUNS number.
\HETH={93/12}   %HE/TH number.
%\monthyear={February, 2010} %If you want to fix the date.

%%%%%%%%%%%%%%%%%%%%% Macros %%%%%%%%%%%%%%%%%%%%%%%%%
\def\meson{\ket{B({\bf q})}}
\def\int{\intop\nolimits}

\def\q{{\bf q}}
\def\calA{{\cal A}}
\def\calM{{\cal M}}
\def\E{{\rm E}}
\def\k{k_1}
\def\u{{u'}}
\def\x{{x'}}
\def\sqtt{\sqrt{t^2-1}}
\def\op#1#2{#1 \otimes #2}
\def\vsl{v \kern -5.4pt / }
\def\psl{p \kern -5.2pt / }
\def\ppsl{\mib{p} \llap / }
\def\epsl{\epsilon \kern -5.0pt /}
\def\qsl{q \llap /}
\def\SL{S_{\rm L}}
\def\SLz{S_{\rm L0}}
\def\SLo{S_{\rm L1}}
\def\SLt{S_{\rm L2}}
\def\SH{S_{\rm H}}
\def\SHm{S_{{\rm H}-1}}
\def\SHz{S_{\rm H0}}
\def\SHo{S_{\rm H1}}

\def\mM{\big({m \over M}\big)}
\def\Iopk{I_1^{{\bf p}\cdot{\bf k}}}
\def\Itpk{I_2^{{\bf p}\cdot{\bf k}}}

\def\mymatrix#1#2#3#4{\left( \matrix{#1  &  #2  \cr
                                      #3  &  #4  \cr} \right) }
\def\myvector#1#2{\pmatrix{#1 \cr #2  \cr}}
\def\colvector#1#2{\big( #1 \ \  #2 \big) }
\def\drangle{\rangle\kern-2.8pt\rangle}
\def\dlangle{\langle\kern-2.8pt\langle}
\def\dvert{\vert\kern-2pt\vert}
\def\dbra#1{\dlangle #1 \dvert}
\def\dket#1{\dvert #1 \drangle}
%%%%%%%%%%%%%%%%%%%%%%%%%%%%%%%%%%%%%%%%%%%%%%%%%%%%%%
\def\NGL{N_{\rm GL}}
\def\LBS{\Lambda_{\rm qcd}}
\def\FPS{F_\pi({\rm PS})}

\def\FB{F_B}

\def\caption#1{\centerline{\vbox{#1}}}
\def\frac#1#2{{#1 \over #2}}
\def\figitem#1{\item{}}
\def\draw#1#2{\vbox{\vbox{\centerline {#1}}\vbox{#2}}}
%%%%%%%%%%% References %%%%%%%%%%%%%%%%%%%%%%%%%%%%%%%%%%%%%%%%%
\REF\IWtwo{
	N. Isgur and M.B. Wise
	\journal Phys. Rev. Lett. &66 (91) 1130.
}
\REF\IW{
	N. Isgur and M.B. Wise
	\journal Phys. Lett. &B232 (89) 113;
	\andjournal Phys. Lett. &B237 (90) 527.
}
\REF\EH{
	E. Eichten and B. Hill
	\journal Phys. Lett. &234 (89) 511.
}
\REF\HG{
	H. Georgi
	\journal Phys. Lett. &240 (90) 447.
}
\REF\KM{
	M. Kobayashi and T. Maskawa \journal
	Prog. Theor. Phys. &49 (73) 652.
}
%%%
\REF\ABKMN{
	K-I. Aoki, M. Bando, T. Kugo, M.G. Mitchard and
	H. Nakatani \journal Prog. Theor. Phys. &84 (90) 683.
}
\REF\AKM{
	K-I. Aoki, T. Kugo and M.G. Mitchard \journal
	Phys. Lett. &B266 (91) 467.
}
\REF\NAKANISHI{
	See for example N.~Nakanishi
	\journal Prog. Theor. Phys. Suppl. &43 (69) 1.
}
\REF\ASYMPTOTICS{
	K-I. Aoki, M. Bando, T. Kugo and M.G. Mitchard
	\journal Prog. Theor. Phys. &85 (91) 355.
}
%%%
\REF\JM{
	P. Jain and H.J. Munczek \journal
	Phys. Rev. &D44 (91) 1873; \andjournal
	Phys. Rev. &D46 (92) 438.
}
\REF\KuMiWT{
	T. Kugo and M.G. Mitchard \journal
	Phys. Lett. &B282 (92) 162.
}
\REF\GP{
	H. Georgi and H.D. Politzer \journal
	Phys. Rev. &D14 (76) 1829.
}
\REF\PS{
	H. Pagels and S. Stokar \journal
	Phys. Rev. &D20 (79) 2947.
}
%%%
\REF\KuMi{
	T. Kugo and M.G. Mitchard \journal
	Phys. Lett. &B286 (92) 355.
}
\REF\GL{
	J. Gasser and H. Leutwyler \journal
	Phys. Rep. &87 (82) 77.
}
\REF\PDG{
	Particle Data Group \journal
	Phys. Rev. &D45 (92) S1.
}
\REF\Rosner{
	J.L. Rosner \journal
	Phys. Rev. &D42 (90) 3732.
}
\REF\Suzuki{
	M. Suzuki \journal
	Phys. Lett. &162B (85) 392.
}
\REF\JHD{
	H.-Y. Jin, C.-S. Huang and Y.-B. Dai \journal
	Z. Phys. &C56 (92) 707.
}
%%%
\REF\MRR{
	T. Mannel, W. Roberts and Z. Ryzak \journal
	Phys. Lett. &B254 (91) 274.
}
\REF\PB{
	P. Ball, preprint HD-THEP-92-25,
        to appear in  Proceedings of ``27th Rencotres de Moriond'',  1992.
}
\REF\KS{
	J.G. K\"orner and G.A. Schuler \journal
	Z. Phys. &C38 (88) 511.
}
\REF\BSW{ % I-W func from Oscillator model
	M. Bauer, B. Stech and M. Wibel \journal
	Z. Phys. &C29 (85) 637.
}
\REF\GISW{
	B. Grinstein, N. Isgur, D. Scora and M.B. Wise \journal
	Phys. Rev. &D39 (89) 799.
}
%%%
\REF\HMW{
	K. Hagiwara, A.D. Martin and M.F. Wade \journal
	Nucl. Phys. &B327 (89) 569.
}
\REF\CKP{
	J.M. Cline, G. Kramer and W.F. Palmer \journal
	Phys. Rev. &D40 (89) 793.
}
\REF\ARGUS{
	ARGUS Collaboration \journal
	Z. Phys. &C57 (93) 533.
}
\REF\NR{ % charge radius from Oscillator model
	M. Neubert and V. Rieckert \journal
	Nucl. Phys. &B382 (92) 97.
}
\REF\Neu{ % QCD sum rule
	M. Neubert \journal Phys. Rev. &D45 (92) 2451.
}
%%%
\REF\Neub{ %  Model-independent extraction of Vcb ...
	M. Neubert \journal
	Phys. Lett. &B264 (91) 455.
}

%%%%%%%%%%%%%%%%%%%%%%%%%%%%%%%%%%%%%%%%%%%%%%%%%%%%%%%%%%%%%
\titlepage
\title{Isgur-Wise Function From Bethe-Salpeter Amplitude}
\author{Taichiro KUGO,  Mark G.~MITCHARD  and Yuhsuke YOSHIDA}
\Kyoto
\abstract{
We develop the improved ladder approximation to QCD in order
to apply it to the heavy quark mesons. The resulting
Bethe-Salpeter equation is expanded in powers of the inverse heavy
quark mass $1/M$, and is shown to be consistent with the heavy quark spin
symmetry.
We calculate numerically the universal
leading order BS amplitude for heavy pseudoscalar
and vector mesons, and use this to evaluate the Isgur-Wise function and
the decay constant $F_B$.
The resulting Isgur-Wise function predicts a large charge radius,
$\rho ^2 = 1.8 - 2.0$, which when fitted to the ARGUS data
corresponds to the value
$V_{\rm cb} = .044 - .050$ for the Kobayashi-Maskawa matrix
element.
}
\endpage

\chapter{Introduction}

Recently there has been a great deal of interest in using $B$ meson
physics to test the detailed structure of the Standard Model and to
reveal new physics.
When we consider systems containing a heavy quark (such as the $b$
quark and, possibly, the $c$ quark) a new symmetry, the heavy quark
spin-flavour symmetry\refmark{\IWtwo} appears.
This symmetry, which becomes exact in the limit when the heavy quark
mass $M$ goes to infinity, may be used to predict various properties
of the heavy mesons; these have been systematically studied using the
Heavy Quark Effective Theory.\refmark{\IW,\EH,\HG}
For instance, when $M\rightarrow  \infty $, every semi-leptonic form
factor can be expressed in terms of a single universal function, the
Isgur-Wise function.\refmark{\IW}
At the moment there is a great deal of work being done to understand
the semi-leptonic weak decay process $B\rightarrow D^{(*)}l\nu$ and
thence to extract the Kobayashi-Maskawa matrix element $V_{\rm
cb}$:\refmark{\KM} it is therefore important to have a reliable
functional form for the Isgur-Wise function.
Unfortunately the heavy quark symmetry tells us nothing about the
Isgur-Wise function away from the kinematical end point.
In order to calculate it at any other point we have to know the
details of the strong interaction physics of the light degrees of
freedom in the heavy mesons.

In this paper we develop the improved ladder approximation to the QCD
Bethe-Salpeter equation as an expansion in powers of $1/M$.
We show that it is indeed consistent with the heavy quark symmetry,
and apply the results to calculate the Isgur-Wise function.
In previous papers\refmark{\ABKMN,\AKM} we have shown that the
improved ladder approximation is capable of giving rather a good
description of the physics of light mesons, and we therefore expect it
to provide a reasonable approximation to the physics of the light
degrees of freedom in heavy quark mesons.

We calculate the Isgur-Wise function as a one loop diagram containing
two meson BS amplitudes and a single current insertion.
The main problem which we must handle is the following: whilst the
overall boundstate  momenta must remain timelike, we have to
Wick-rotate the loop momentum to evaluate the integral.
Then the combinations of the loop momentum and the overall boundstate
momentum which appear in the arguments of the quark mass functions
(and also, depending on the improvement scheme, in the argument of the
running coupling constant) becomes complex (non-real).
Moreover, when we calculate the Isgur-Wise function, we work in the
rest frame of one of the two mesons but the other must be boosted to a
finite velocity, and this corresponds to the BS amplitude with complex
relative momentum.
We are thus forced to calculate the BS amplitude with complex
arguments.
We propose a new method to do this analytic continuation using the BS
equation itself.
Because of this complication, the numerical work becomes quite hard
--- the Isgur-Wise function, for instance, appears as a
five-dimensional integral.

In order to illustrate the main points of the method and obtain
numerical results without being too much involved in such
complications, we here ignore the running of the quark masses and work
with a fixed value of the light quark mass $m$.
We determine $m$ by solving the equation $m=\Sigma(am^2)$ where
$\Sigma$ is the light-quark mass function determined from the ladder
Schwinger-Dyson equation, and $a$ is an unknown parameter.
We have calculated the Isgur-Wise function for the two cases which we
regard as typical, $a=1$ and $a=4$.
A full treatment including the quark mass functions derived from the SD
equation will be reserved for a  later work.
We actually find that the general shape of the Isgur-Wise function and
the value of the charge radius $\rho^2$ are rather independent of the
value of the fixed light-quark mass.
Consequently we expect that the `true' values obtained by including
the mass function $\Sigma(x)$ will not differ significantly from our
present results.
The heavy meson decay constants, on the other hand, which we also
calculate do not show this independence of the value of $m$, and so we
have to wait for the full treatment.

The rest of this paper is organised as follows.
In section 2 we present our formalism for treating the heavy meson BS
equation as an expansion in powers of inverse heavy quark mass $1/M$,
and we derive the leading order BS equation in a component form.
An exact expression for the decay constant $F_B$ in terms of the BS
amplitude is given in section 3.
In section 4 we explain how to calculate the Isgur-Wise function and
show that our approximation is actually consistent with current
conservation in the equal heavy quark mass case.
In section 5 we demonstrate that the leading order BS equation
satisfies the heavy quark spin symmetry and discuss the implication
for the form of BS amplitudes for pseudoscalar, vector, scalar and
axial-vector mesons.
Section 6 is devoted to the details of the numerical calculations.
In section 6.1 we fix a suitable analytic functional form for the
running coupling constant.
In section 6.2 we use this coupling and solve the SD equation for the
light quark mass function $\Sigma(x)$.
We need $\Sigma$ both in order to fix the light quark mass which we
use in the BS equation, and also to fix the overall mass scale in MeV.
We fix the scale by calculating the Pagels-Stokar pion decay constant
and defining the result to $93\sqrt{2}$MeV.
In section 6.3 we solve the BS equation and calculate the decay
constant $F_B$ and the energy eigenvalues of the ground state and the
first excited state.
Finally, in section 6.4, we present the results of our calculation of
the Isgur-Wise function $\xi(t)$.

\chapter{BS Amplitude and Notation}

The BS amplitude of a meson boundstate $\meson$ of total momentum
$q$ containing a heavy quark
$\Psi $ and a light anti-quark $\bar \psi $ is defined by
$$
\bra{0} {\rm T} \Psi_{i\alpha }(x)
\bar\psi_\beta ^j(y) \ket{B(\q)}
= e^{-iqX}\int {d^4p\over (2\pi )^4}\,e^{-ipr}\
\delta _i^j\,\chi _{\alpha \beta }(p;q) \ ,
\eqn\eqBSAMP
$$
where $i, j$ and $\alpha , \beta $
denote colour and Dirac spinor indices respectively.
The coordinate $X^\mu $ and the relative coordinate $r^\mu $ are
defined by
$$
X^\mu  \equiv  \zeta x^\mu  + \eta y^\mu  \ , \qquad \
r^\mu  \equiv  x^\mu -y^\mu  \ .
$$
where $\zeta $ and $\eta $ are real parameters satisfying
$\zeta +\eta =1$. $p$ is the therefore the relative momentum.
Although in the relativistic case any value may in principle
be chosen for $\zeta $, we make the following `natural' choice
$$
\zeta ={M\over M+m}\ , \ \qquad \ \eta ={m\over M+m}\ ,
$$
with $M$ and $m$ being the masses of the heavy and light quarks,
respectively. With this choice, $X^\mu $ is the centre-of-mass
coordinate of the system; it has the technical advantage of
making it legitimate to perform a Wick rotation in the
BS equation.\refmark{\NAKANISHI}
For later use we here define the 4-velocity
$v^\mu $ of the boundstate:
$$
v^\mu  = q^\mu  / M_B\ , \quad \qquad  v^2 = 1 \ ,
\ee$$
where $M_B$ is the boundstate mass, $q^2=M_B^2$.

It is convenient to introduce a ket notation to denote
the bispinor BS amplitude:
$$
\dket{\chi } \ \  \longleftrightarrow \ \
\delta _i^j\, \chi _{\alpha \beta }(p;q) \ .
\ee
$$
The conjugate BS amplitude which is denoted by a bra state
is defined by
$$
\dbra{\chi } \ \  \longleftrightarrow \ \
\delta _i^j\, \bar\chi _{\alpha \beta }(p;q)
\equiv  \delta _i^j\, \gamma _0 \,
[\chi _{\alpha \beta }(p^*;q)]^{\dagger}\gamma _0 \ .
\eqn\eqCONJG
$$
Note that the complex conjugate $p^*$ of the argument $p$
appears in the RHS: this is necessary for the general case
where the relative momentum $p$
becomes complex after Wick rotation.
%, and corresponds to the
%completeness relation $\int _p \dket{p^*}\dbra{p}=1$.
In fact, even when $p$ is real, the $i\epsilon $
prescription in the Feynman propagator implies
that the Minkowskian theory should be understood
as that obtained by analytic continuation of the
Euclidian theory. Considering the limit
${\rm Arg}\,p^0 \rightarrow  0+$,
one can see that $\bar\chi $ defined this way does
indeed coincide with the conventional conjugate BS
amplitude:\foot{One can also confirm this fact by
seeing that the BS equation for the conventional
conjugate BS amplitude is indeed satisfied
by our conjugate amplitude \eqCONJG.}
$$
\bra{B(\q)} {\rm T} \psi _{j\beta }(y)
\bar\Psi _\alpha ^i(x) \ket{0}
= e^{iqX}\int {d^4p\over (2\pi )^4}\,e^{ipr}\
\delta _j^i\,\bar\chi _{\beta \alpha }(p;q) \ .
\eqn\eqCONJGBSAMP
$$
We also introduce a (non-positive-definite) inner product:
$$
\dbra{\psi }\chi \drangle \equiv
N_c \int {d^4p\over i(2\pi )^4}
\tr[\bar\psi (p;q) \chi (p;q)] \ ,
\eqn\eqINNERPRODUCT
$$
where $N_c=3$ comes from the trace over the colour index
and the trace in the RHS
is taken over the bispinor indices alone.

The bispinor BS amplitude $\chi $ may be expanded into
invariant component amplitudes.
Here we are mainly interested in
the pseudoscalar boundstate for which
here there are four such amplitudes:
in the present context of the heavy quark system,
it is convenient to use the form
$$
\chi (p;q)\  =
  \Lambda _+\big( A(u,x) + B(u,x)\ppsl \big) \gamma _5
  + \Lambda _-\big( C(u,x) + D(u,x)\ppsl \big) \gamma _5\ ,
\ee
$$
where the projection operator $\Lambda _{\pm }$ and
the 4-momentum ${\mib p}^\mu $, the spatial projection of $p^\mu$,
are defined by
$$
\Lambda _{\pm } \equiv  {1\pm \vsl\over 2}\ ,  \qquad \quad
{\mib p}^\mu  \equiv  p^\mu  - (p\cdot v)v^\mu \ ,
$$
and the variables $u$ and $x$ are
$$
iu = p\cdot v\ , \quad \qquad x=\sqrt{-{\mib p}^2}
=\sqrt{-p^2-u^2} \ .
\ee
$$
The components of the conjugate BS amplitude $\bar\chi $ are
defined similarly:
$$
\bar\chi (p) =
- \gamma _5\big( \bar A(u,x)
  + \bar B(u,x)\ppsl \big) \Lambda _+
- \gamma _5\big( \bar C(u,x)
  + \bar D(u,x)\ppsl \big) \Lambda _- \ .
\ee
$$
Note that $iu$ is the time component $p^0$ of $p^\mu $
in the restframe of the boundstate where
$v^\mu =(1,{\bf 0})$. In this frame, $u$ becomes real
after Wick rotation,
$p^\mu  \ \rightarrow \ p^\mu _{\rm E}=(u,{\bf p})$,
and then eq.\eqCONJG\ gives
$$
\bar\chi (p) =
-\gamma _5\big( A^*(-u,x)
  + B^*(-u,x)\ppsl \big) \Lambda _+
- \gamma _5\big( C^*(-u,x)
  + D^*(-u,x)\ppsl \big) \Lambda _-  \ .
\ee
$$
In view of the BS equation (given shortly in (2.13)), we can easily
see the relations $X^*(-u,x) \nextline = X(u,x)$ for $X = A, B, C, D$,
{\it provided that} the boundstate mass $M_B$,
determined as a eigenvalue of the BS equation, is {\it real}.
This is because when $q^\mu =M_Bv^\mu $ is real,
in the Wick rotated form of the BS equation
(in which the integration factor $\int d^4k/i(2\pi )^4$ becomes real),
the imaginary quantity appearing is only through $p^0=iu$
(or $k_0\equiv iv$) so that the complex conjugation of the whole equation
is equivalent to changing the sign of $u$ (and $v$).
So, for the true boundstate solutions corresponding to real mass
eigenvalue $M_B$, we have simply
$$
\eqalign{
&\bar A(u,x) =  A(u,x) \qquad \bar B(u,x) =  B(u,x) \cr
&\bar C(u,x) =  C(u,x) \qquad \bar D(u,x) =  D(u,x) \ .\cr
}\ee
$$

To simplify the discussion, we use the following
abbreviation for momentum integration
and for gamma matrix multiplication on the bispinor $\chi$:
$$
\eqalign{
\int _p \equiv  \int {d^4p\over i(2\pi )^4}
&\equiv  \int {d^4p_{\rm E}\over (2\pi )^4}
\equiv  \int {x^2dx \, du \, d\!\cos \theta \over 8\pi ^3} \ ,\cr
\Gamma  \otimes \Delta \ \dket{\chi } \ \ &\longleftrightarrow \ \
\delta _i^j\ (\Gamma \,\chi \,\Delta )_{\alpha \beta } \ .\cr
}\ee
$$
In this notation the BS equation becomes
$$
\op{\SH(p+\zeta q)}{\SL(p-\eta q)}\,\dket{\chi } = K \dket{\chi }
\eqn\eqFULLBS
$$
where $\SH$ and $\SL$ are ($i$ times) the inverse propagators
of the heavy quark and the light quark respectively
$$
\SH \equiv  \psl + \zeta \qsl -M \ ,\qquad \quad
\SL \equiv  \psl - \eta \qsl -m \ ,
\ee
$$
and $K$ stands for the Bethe-Salpeter kernel.
In the improved ladder approximation $K$ is given by the
following one-gluon-exchange form:
$$
\eqalign{
K\dket{\chi }
&\equiv  \int _k g^2 C_2\,D_{\mu \nu }(p-k)
  \gamma^\mu  \chi (k) \gamma^\nu  \ ,\cr
&D_{\mu \nu }(k) =
  {g_{\mu \nu }-k_\mu k_\nu /k^2 \over k^2} \ ,\cr
}\ee
$$
where $C_2$ is the second Casimir given by
$C_2=(N_c^2-1)/2N_c$ and $g^2$ is a suitable
running coupling constant. Since the running
of the coupling is mainly caused by the gluon
self-energy corrections, probably the
best choice of the argument of the running coupling constant
is to use the gluon momentum and take $g^2\big((p_\E-k_\E)^2\big)$,
where $p_\E$ is Euclidean momentum; $p_\E^2 = - p^2$.
In this paper we take for simplicity
$g^2\big(p_\E^2+k_\E^2\big)$.
We have discussed this choice of argument before\refmark{\ASYMPTOTICS}
 and have shown that it does not differ from the
ideal case very much. The precise functional form of
our running coupling constant will be given later.

In the context of the heavy quark system, it is convenient to adopt
the following boundstate normalisation
$$
\VEV{B(\q)\vert B(\q')} = (2\pi )^3 \, 2v_0 \, \delta ^3(\q-\q')\ .
\ee
$$
This differs from the usual invariant normalisation
in that we have used $v^0$ in place of $q^0=M_B v^0$:
the usual invariantly normalised state is therefore given by
$\ket{B({\bf q})}\sqrt{M_B}$.
By considering the contribution of the intermediate boundstate
to the two heavy two light quark Greens function in the
standard way, we can show that our BS amplitude may be
normalised by requiring\refmark{\NAKANISHI}
$$
\zeta  \dbra{\chi } \op{\gamma _\mu }{\SL} \dket{\chi }
-\eta  \dbra{\chi } \op{\SH}{\gamma _\mu } \dket{\chi }
= 2 v_\mu  \ .
\eqn\eqNORM
$$

The BS equation gives an eigenvalue equation for the
boundstate mass $M_B$ or, equivalently, for the binding energy $E$
defined as
$$
M_B=M+m-E = \zeta ^{-1}( M - \zeta E ) \ .
\ee
$$

We now expand all quantities as power series in $m/M$.
The light quark mass $m$ (which
is the constituent quark mass and comes almost entirely from dynamical
chiral symmetry breaking) is of order $\Lambda _{\rm QCD}$,
and we therefore regard $\Lambda _{\rm QCD}$ as $O(1)$
in this expansion.
Thanks to our choice of $\zeta $ and $\eta $,
the relative momenta $p$ and $k$ are also of the order of
$\Lambda _{\rm QCD}$, and therefore also $O(1)$.
The binding energy and
BS amplitude are expanded as
$$
\eqalign{
&\zeta E=E_0+\mM E_1 +\mM^2 E_2 + \cdots  \ ,\cr
&\chi =\chi _0 + \mM \chi _1 + \mM^2 \chi _2 + \cdots \ , \cr
}\ee
$$
with corresponding invariant amplitudes:
$$
\eqalign{
&\chi _0(p) = \Lambda _+\big( A_0(u,x)
  + B_0(u,x)\ppsl \big) \gamma _5
+ \Lambda _-\big( C_0(u,x)
  + D_0(u,x)\ppsl \big) \gamma _5 \ ,\cr
&\chi _1(p) = \Lambda _+\big( A_1(u,x)
  + B_1(u,x)\ppsl \big) \gamma _5
+ \Lambda _-\big( C_1(u,x)
  + D_1(u,x)\ppsl \big) \gamma _5 \ ,\cr
}\ee
$$
and so on.
The expansion of the quark inverse propagators is (notice
that $\SH$ starts with a $-1$-st order term)
$$
\eqalign{
\SH &= \mM^{-1}\SHm + \SHz + \mM\SHo + \cdots  \ ,\cr
&\SHm \equiv  m(\vsl -1) = -2m\Lambda _-  \ ,\cr
&\SHz \equiv  \psl - E_0\vsl  \ ,\cr
&\SHo \equiv  -E_1 \vsl \ ,\cr
\SL &= \SLz + \mM\SLo + \mM^2\SLt + \cdots  \ ,\cr
&\SLz \equiv  \psl - m(1+\vsl) = \psl - 2m\Lambda _+  \ ,\cr
&\SLo \equiv  E_0 \vsl \ ,\cr
&\SLt \equiv  E_1 \vsl \ .\cr
}\ee
$$
Substituting these expansions into the BS equation \eqFULLBS,
we find that at $-1$-st order
$$
\op{\SHm}{\SLz} \dket{\chi _0} = 0 \ ,
\eqn\eqMFIRST
$$
at zeroth order
$$
\SHm \otimes
\Big( \SLz \dket{\chi _1} + \SLo \dket{\chi _0} \Big)
+ \op{\SHz}{\SLz} \dket{\chi _0} = K \dket{\chi _0} \ ,
\eqn\eqZEROTH
$$
and so on. Since $\SHm \propto \Lambda _-$,
the $-1$-st order equation \eqMFIRST\ simply
implies that $\chi _0 \propto \Lambda _+ $ so that
we find $C_0(u,x)=D_0(u,x)=0$.
This then causes the term $\op{\SHm}{\SLo} \dket{\chi _0}$
in the zeroth order equation to vanish.
Moreover in the zeroth order equation,
the first term, $\op{\SHm}{\SLz} \dket{\chi _1}$, contributes
only to the piece proportional to $\Lambda _-$, so that we have
$$
 \Lambda _+ \big[ \op{\SHz}{\SLz} -K \big] \dket{\chi _0} = 0 \ .
\eqn\eqBSZERO
$$
Projecting out the $\Lambda _+$ pieces by taking
${1\over 2}\tr[\gamma _5\Lambda _+\times  $\eqBSZERO$]$ and
${1\over 2}\tr[\gamma _5\ppsl \Lambda _+\times  $\eqBSZERO$]$,
the zeroth (or leading) order BS equation \eqBSZERO\
becomes
$$
(E_0-iu) {\cal M} \myvector{A_0(p)}{B_0(p)}
=\int {y^2dydv\over 4\pi ^3}\, g^2C_2\,K_0(p,k)
\myvector{A_0(k)}{B_0(k)} \ ,
\eqn\eqZEROC
$$
where the arguments $p$ and $k$ stand for $(u,x)$ and
$(v,y)$ respectively, $\calM$ is a `metric' matrix given by
$$
\calM \equiv  \mymatrix{iu}{-x^2}{-x^2}{x^2(iu-2m)}\ ,
\eqn\eqMATRIXM$$
and the kernel $K_0(p,k)$ is obtained after
integrating $K$ over the angular variable
$\cos\theta = {\bf k}\cdot {\bf p}/\abs{\bf k}\abs{\bf p}$.
The explicit matrix expression for $K_0(p,k)$
is given in the Appendix.

In the leading order of the $m/M$ expansion
the normalisation condition \eqNORM\ becomes
$$
{1\over 2}\dbra{\chi _0} \op{\vsl}{\SLz} \dket{\chi _0} =1\ ,
\eqn\eqNORMZ
$$
which when rewritten in terms of the invariant amplitudes gives
$$
N_c\int {x^2dxdu\over 4\pi ^3}\,
\left\{ \colvector{\bar A_0(p)}{\bar B_0(p)}
\calM \myvector{A_0(p)}{B_0(p)} \right\} = 1
\eqn\eqTHENORM
$$
with the metric matrix $\calM$ defined in eq.\eqMATRIXM.

The BS equation \eqZEROC\ is a linear eigenvalue equation
$\calA \,\dket{\chi}= E_0 \calM\,\dket{\chi}$
with $\calA=g^2C_2K_0+iu\calM$ and metric $\calM$.
Further, both $\calA$ and the metric $\calM$ are hermitian
with respect to the inner product $\dlangle{\psi}\dket{\chi}$ defined
in \eqINNERPRODUCT:
$[\calA_{i,j}(-u,x;-v,y)]^*=\calA_{j,i}(v,y;u,x)\ (i,j=1,2)$
and so on. As suggested by the normalisation condition (2.28),
the natural inner product to this BS system is given by
$\dbra{\psi}{\cal M}\dket{\chi}$ and the norm by
$\dvert \chi\dvert^2=\dbra{\chi}{\cal M}\dket{\chi}$ accordingly.
Indeed, the BS equation \eqZEROC, with the help of the hermiticity of
${\cal A}$ and ${\cal M}$, leads to an equality
$(E_0^{\psi*}-E_0^{\chi}) \dbra{\psi}{\cal M}\dket{\chi} = 0$ for
arbitrary two eigenstates $\dket{\chi}$ and $\dket{\psi}$ belonging to
eigenvalues $E_0^{\chi}$ and $E_0^{\psi}$,respectively. This implies,
in particular, that the energy eigenvalues $E_0$ must be {\it real}
as far as the eigenstate has non-zero norm
$\dbra{\chi}{\cal M}\dket{\chi}\not=0$. Namely
all solutions to our BS equation \eqZEROC\ for which the LHS of
Eq.(2.28) is non-zero correspond to real eigenvalues $E_0$.
In numerical work we actually find
many solutions to \eqZEROC\ with complex eigenvalues $E_0$,
but they all turn out to have zero-norm in
accord with this observation.

\chapter{Decay Constant}

The pseudoscalar meson decay constant $F_B$ is defined by
$$
iF_B q_\mu  e^{-iqx} \equiv
\bra{0} \bar\psi (x)\gamma _\mu \gamma _5\Psi (x)
\ket{B({\bf q})}\sqrt{M_B} \ ,
\eqn\eqFBDEF
$$
where the factor of $\sqrt{M_B}$ is included so that the state
$\ket{B({\bf q})}\sqrt{M_B}$ satisfies the usual relativistic
normalisation.
Our definition would correspond to a pion decay constant
$F_\pi=93\times\sqrt2$ MeV.

Setting $r\equiv x-y=0$ in the definition of the BS
amplitude \eqBSAMP\ we find
$$
F_B\, q_\mu  = - N_c\int _p
\tr[ \gamma _\mu \gamma _5 \chi (p;q) ] \times \sqrt{M_B}\ .
\ee
$$
Substituting the component expression of the BS
amplitude $\chi $, we have
$$
F_B\sqrt{M_B} = N_c \int {x^2dxdu\over 2\pi ^3} \,
	[\ A(u,x) - C(u,x)\ ] \ ,
\eqn\eqFB
$$
which is an exact formula.

\chapter{Isgur-Wise Function}

The expression for a heavy quark current operator
$\bar\Psi '\gamma _\mu \Psi $
between two pseudoscalar boundstates $B'$ and $B$ containing
heavy quarks $\Psi '$ and $\Psi $ with masses $M'$ and $M$
may be described in terms of two form factors $f(t)$ and
$g(t)$ as
$$
\bra{B'(\q')}\bar\Psi '(x)\gamma _\mu \Psi (x)\ket{B(\q)} \equiv
\big[ (v+v')_\mu  f(t) + (v-v')_\mu  g(t) \big] e^{i(q'-q)x}
\eqn\eqSANITI
$$
where $t = v\cdot v'$.
When the initial and final mesons are the same (that is $\Psi =\Psi '$
and $M_B=M_{B'}$), current conservation leads to $g(t)=0$.
The Isgur-Wise function $\xi (t)$\refmark{\IW} is defined as the leading
term in the expansion of the first
form factor $f(t)$ in powers of $m/M$ (or $m/M'$)
$$
\eqalign{
f(t) &= \xi (t) + \mM f_1(t) +\mM^2 f_2(t) + \cdots \ ,\cr
g(t) &=  \mM g_1(t) +\mM^2 g_2(t) + \cdots \ .\cr
}\ee
$$
The absence of the leading order term $g_0(t)$ in $g(t)$ can easily be
seen :  using the equality $\Lambda_+(\vsl'-\vsl)\Lambda_+=0$, we find
$ \bar\chi_0 (p';q')(\vsl'-\vsl) \chi_0 (p;q) = 0$
(remember $\chi_0 = \Lambda_+\chi_0$ and
$\bar \chi_0 = \bar \chi_0\Lambda_+$).

$\xi (t)$  is independent of the heavy quark masses
$M$ and $M'$ and is universal over pseudoscalar and vector
heavy-light quarkonia: exactly the same function appears
in the expansion of the form factors in the vector meson case.
\FIG\FIGISGWIS
	%{Figure \FIGISGWIS. Feynman diagram contributing to
	%	the Isgur-Wise function $\xi(t)$..}
In the present approximation using constant mass for the heavy quark,
the vector current matrix element \eqSANITI\ may be calculated
from the diagram of figure \FIGISGWIS.
\vskip10pt
\centerline{\epsfbox{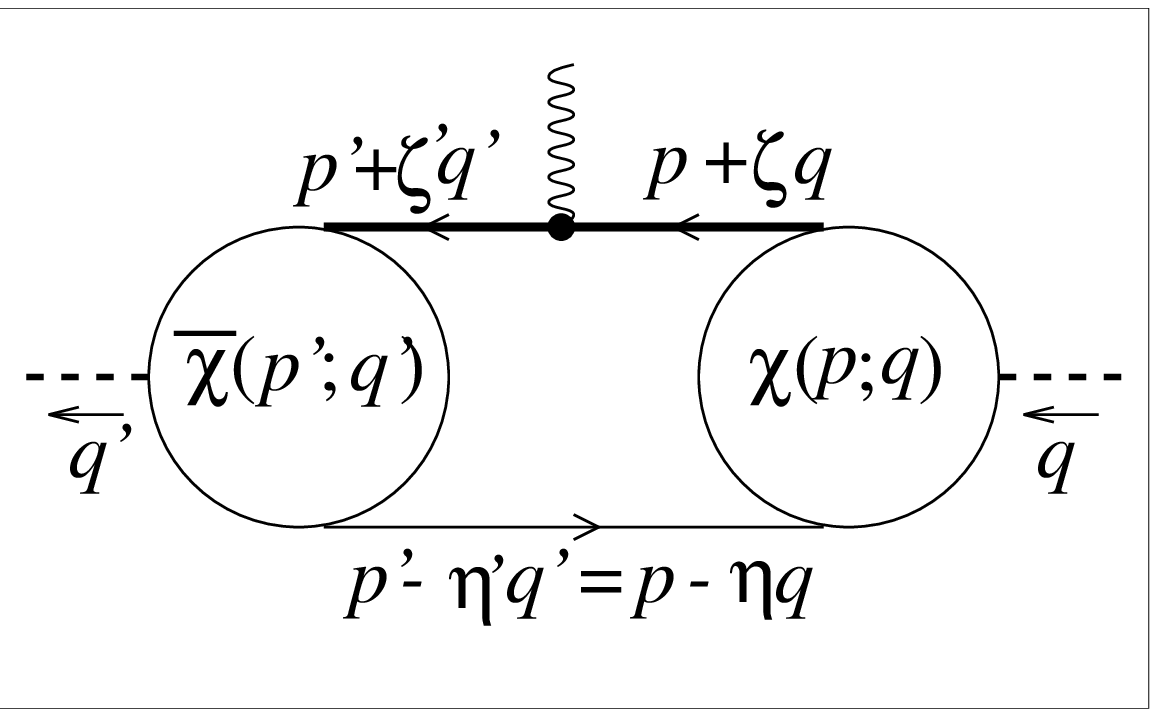}}
\centerline{Fig.\FIGISGWIS\ Feynman diagram used to calculate the
Isgur-Wise function.}
\vskip10pt
\noindent
Thus we use
$$
\eqalign{
 \bra{B'(\q')}\bar\Psi '(0)&\gamma _\mu \Psi (0)\ket{B(\q)} \cr
 &= N_c \int {d^4p'_{\rm E}\over (2\pi )^4}
\tr\big[ \bar\chi (p';q')\gamma _\mu \chi (p;q)
\SL(p-\eta q =p'-\eta 'q') \big] \ ,\cr
}\eqn\eqLADDERISGWISE
$$
where $\eta '\equiv m/(M'+m)$ and $\zeta '\equiv M'/(M'+m)$
for the final state meson $B'$.
Actually, in the equal mass case, the approximation of eq.\eqLADDERISGWISE\
is consistent with current conservation to all orders:
using $\eta +\zeta =\eta '+\zeta '=1$, $p-p'=\eta q-\eta 'q'$
and $M=M'$ we have
$$
\eqalign{
q-q' &= (\eta +\zeta )q-(\eta '+\zeta ')q'
  =(p+\zeta q)-(p'+\zeta 'q') \cr
\longrightarrow& \ \ \qsl-\qsl'
  = (\psl+\zeta \qsl-M ) - (\psl'+\zeta '\qsl'-M )
  = \SH - \SH' \ .
}\ee
$$
Using this equation and the BS equation \eqFULLBS, it is easy
to show that $g(t) = 0$:
$$
\eqalign{
M_B (v-v')^2 g(t) &=
(q-q')^\mu \dbra{\chi '} \op{\gamma _\mu }{\SL}\dket{\chi } \cr
&=\dbra{\chi '} \op{(\qsl-\qsl')}{\SL}\dket{\chi } \cr
&=\dbra{\chi '} \big(\op{\SH}{\SL}-\op{\SH'}{\SL'}\big)\dket{\chi }
\qquad ({\rm since\ } \SL=\SL' ) \cr
&=\dbra{\chi '} (\ K\ - \ K\ ) \dket{\chi } = 0 \ . \cr
}\ee
$$

We make one small comment:
it is known that if a running heavy quark mass function determined from the
ladder SD equation were used,
the matrix element \eqSANITI\  should be calculated by
including gluon ladders also in the vector current channel,
since otherwise current conservation would be violated.
However diagrams containing one or more gluon exchanges
in this channel are easily seen to be suppressed by at least
a power of $\mM^2$ and so the approximation we use
satisfies current conservation to at least leading order in
the $m/M$ expansion. In fact $g_0(t)=0$ as we have seen before.

The Isgur-Wise function $\xi (t)$ is easily obtained by taking the
leading order term of eq.\eqLADDERISGWISE\ and contracting with
$(v+v')^\mu $: we find
$$
2(1+t)\xi (t) = N_c \int _{p'}\tr\big[
\bar\chi _0(p';q')\,(\vsl+\vsl')\chi _0(p;q)\,
\SLz(p-\eta q =p'-\eta 'q') \big] \ .
\eqn\EQIW
$$
After taking the trace, we find the expression in terms of
the invariant amplitudes:
$$
\xi (t) =  {N_c \over 1+t} \int
{du' \, x'^2 dx' d\cos\theta \over 8\pi^2}\,
\left\{
\colvector{\bar A'_0}{\bar B'_0}
\mymatrix{L_{11}}{L_{12}}{L_{21}}{L_{22}}
\myvector{A_0}{B_0}
\right\} \ ,
\eqn\eqISGWISINT
$$
where $A'_0\equiv A_0(u',x')$, $A_0\equiv A_0(u,x)$ and so on,
and the elements of the matrix $L$ are given by
$$
\eqalign{
L_{11} &= \k + i\u(t+1) \cr
L_{12} &= -\k^2 -i\u\k (2t+1)
  +(t+1)\big[ m\k + i\u(t-1)(m-i\u) - \x^2 \big] \cr
L_{21} &=  \k(2m-i\u) - \x^2(t+1) \cr
L_{22} &=  -\k^2(2m-i\u) + \k t\x^2  \cr
  &\qquad \
 +(t+1)\big[ \k (i\u-m)(i\u-2m)  +i\u\x^2 t - m\x^2 (t+1) \big] \cr
}\ee
$$
with
$$
\k \equiv  p'\cdot (v - v't) = p'_1\sqtt \ .
\ee
$$
We should note that the important condition $\xi(t\!\!=\!\!1)=1$ is satisfied
in our formulation:
actually, Eq.\EQIW\ may be rewritten into the form
$$
\xi(t) =
\frac{1}{1+t}\dbra{\chi_0'}\frac{\vsl+\vsl'}{2}\otimes\SLz
	\dket{\chi_0}\ .
$$
where $t$ becomes 1, $v'$ and $\chi_0'$ become $v$ and $\chi_0$,
respectively, so that the RHS of this clearly reduces to the
normalisation condition \eqNORMZ\ and gives 1.

We evaluate the integral in \eqISGWISINT\
in the $B'$ restframe in which $v'=(1,{\bf 0})$.
In this frame we label the components of $p'$ as
$$
\eqalign{
&p' = ( iu', p'_1, p'_2, p'_3) = ( iu', p'_1, {\bf p}'_\perp) \ ,\cr
&p'_1=x'\cos\theta\ , \qquad  |{\bf p}'_\perp | = x'\sin\theta \ , \cr
}\ee
$$
and take the integration contour of $\int d^4p'_{\rm E}$
over the real axis of $u'$ and $p'_i$.
The momentum $p$ of
the initial state BS amplitude $\chi (p;q)$
is determined by the relation
$$
p-\eta q = p'-\eta 'q' \quad \rightarrow  \quad
p =  p'-\eta 'q' + \eta q   \ .
\eqn\eqP
$$
and becomes complex. We require the corresponding arguments
$u$ and $x$ for the component amplitudes $A_0$ and $B_0$.
Since $u=p\cdot v/i$ and $x=\sqrt{-p^2+(p\cdot v)^2}$ are
Lorentz invariants, it is simplest to calculate them in
the $v$-restframe where $v=(1,{\bf 0})$.
In this frame the vectors $v'$ and $p'$ take the form
(since $v\cdot v'=t$):
$$
\eqalign{
v'&= (t,\ \sqtt,\ 0, \ 0) \ ,\cr
p'&= (i\u t+p'_1\sqtt, \ \ i\u\sqtt+p'_1t, \ \ {\bf p}'_\perp) \ ,\cr
}\ee
$$
so that $p$ is given by
$$
\eqalign{
p &= p' - \eta' q' + \eta q  = p' - mv' + mv +  O\mM \cr
  &= \Big(i\u t+p'_1\sqtt -m(t-1),\ \ (i\u-m)\sqtt+p'_1t,
\ \ {\bf p}'_\perp \Big) \ + \  O\mM  \ .
}\eqn\eqCOPARA
$$
Therefore to leading order in $m/M$
the arguments $u$ and $x$ of the initial state
BS amplitude $A_0$ and $B_0$ are given by
$$
\eqalign{
u &= {1\over i}p\cdot v = \u t + i[ m(t-1) - p'_1\sqtt ]  \ , \cr
\noalign{\vskip 4mm}
x &= \sqrt{{\bf p}^2}
  = \sqrt{ \big[ i\u\sqtt + (p'_1t-m\sqtt) \big]^2 +
{{\bf p}'_\perp}^2 }  \ .\cr
}\eqn\eqZEROP
$$

Note that $u$ and $x$ in \eqZEROP\  are complex.
In order to find the BS amplitudes $A_0(u,x)$ and $B_0(u,x)$
at these complex values, we need to perform an analytic
continuation. Fortunately this continuation can be done
using the BS equation itself: we simply put
the complex values $u$ and $x$ into the BS equation
\eqZEROC\ and find $A_0(u,x)$ and $B_0(u,x)$ by performing
the RHS integration over $v$ and $y$, which requires only
the knowledge of the BS amplitude with real variables $v$ and $y$.

\chapter{Heavy Quark Spin Symmetry of Leading Order BS Amplitudes}

The full BS equation eq.\eqFULLBS\ of course applies to any
heavy quark light anti-quark boundstate $\meson$.  So the $-1$st order
equation eq.\eqMFIRST\ tells us that the leading order BS amplitude
$\chi _0$ satisfies  $\Lambda _-\chi _0 = 0$ for all boundstates (not
just the pseudoscalar);  $\chi _0\propto \Lambda _+$.
Therefore, for instance, the leading order BS amplitudes of
pseudoscalar and vector boundstates generally have
the following forms:
$$
\eqalign{
\chi _{0 \rm Ps} &= \Lambda _+
  \big( A + B \ppsl\big) \gamma _5 \ ,\cr
\chi _{0 \rm V} &= \Lambda _+ \big[ \epsl( A_V - B_V \ppsl) +
\epsilon \cdot p ( C_V + D_V \ppsl) \big] \ , \cr
}\eqn\eqGENERALBS
$$
where $\epsilon ^\mu $ is the polarisation vector of the
vector boundstate.

Let us now show that our (improved) ladder BS equation satisfies the
heavy quark spin symmetry\refmark{\IW} in the leading order of
$m/M$ expansion, and that the pseudoscalar and vector BS amplitudes
are given by
$$
\eqalign{
\chi _{\rm 0Ps} &=
  \Lambda _+ \gamma _5 \big( A - B \ppsl\big) \ ,\cr
\chi _{\rm 0V} &=
  \Lambda _+  \epsl \big( A - B \ppsl\big ) \ ,\cr
}\eqn\eqPVAMP
$$
with common functions $A$ and $B$, and belong to a common energy
eigenvalue. To see this we rewrite the leading order BS
equation eq.\eqBSZERO\  in the form
$$
(p\cdot v-E_0)\chi _0(p) (\psl - 2m\Lambda _+)
= \int _k v^\mu  \chi _0(k) \gamma ^\nu
\,g^2C_2\,D_{\mu \nu }(p-k)
\eqn\fourpointthree
$$
with $\chi_0$ subject to the constraint $\Lambda_- \chi_0 = 0$.
The important point is that in eq.\fourpointthree\ the heavy
quark spinor index $\alpha $ of the amplitude
${\chi _0}_{\alpha \beta }\,$
(the left-leg)
is left intact: there are no $\gamma $-matrix factors multiplying
$\chi _0$  from the left. The leading order BS equation is
therefore invariant under any heavy quark spin
rotation which commutes with $\Lambda _-$ --- commutativity
with $\Lambda _-$ is required to maintain the constraint
$\Lambda_- \chi_0 = 0$. In particular,
in the restframe of the boundstate in which
$\Lambda _-={1-\gamma _0\over 2}$,
it is invariant under
$$
\eqalign{
\chi _0(p;q) &\rightarrow  U(\bftheta ) \chi _0(p;q) \ , \cr
U(\bftheta )&=
\exp\big(i\bftheta \cdot {\bfsigma\over 2}\big) \ . \cr
}\ee
$$
In this frame, the polarisation vector $\epsilon ^\mu $
of the vector boundstate is of the form
$\epsilon ^\mu =(0,\bfepsilon ) \ (\vert{\bfepsilon }\vert
=1)$
since $\epsilon \cdot v \propto \epsilon \cdot q = 0$.
Let us rotate the pseudoscalar amplitude through $\theta =\pi $
around the $\bfepsilon $-axis by multiplying by
$U(\pi \bfepsilon )$.
Noting the relations
$$
U(\pi \bfepsilon )=i\bfsigma \cdot \bfepsilon  \ \ {\rm and} \ \
\Lambda _+\bfsigma \gamma _5=\Lambda _+\bfgamma \ ,
$$
we find
$$
U(\pi \bfepsilon )\ \Lambda _+ \gamma _5 \big( A - B \ppsl\big) =
i\Lambda _+ \bfepsilon \cdot \bfgamma \big( A - B \ppsl\big) \ .
\ee
$$
But (aside from a constant overall factor $-i$) the RHS
is nothing but the vector meson BS amplitude $\chi_{\rm 0V}$
as given in eq.\eqPVAMP. So if the pseudoscalar BS amplitude
$\chi_{\rm 0Ps}$ in eq.\eqPVAMP\ satisfies the BS equation
\fourpointthree, then so does the vector amplitude $\chi_{\rm 0V}$ in
\eqPVAMP, and the energy eigenvalues are the same.

It is also important to note that the leading order BS equation
\fourpointthree\ is totally independent of the heavy quark mass.
This implies that the excitation energies determined as energy
eigenvalues of that equation are also universal; namely, sufficiently
heavy quark mesons with the same spin-parity or their partners under
heavy quark spin symmetry should show almost the same excitation
energy spectra.

In the same way as in the pseudoscalar and vector case,
we can see that the scalar and axial vector mesons
form a degenerate multiplet under the heavy quark spin symmetry,
and that their leading order BS amplitudes are given in the form.
$$
\eqalign{
\chi _{\rm 0S} &= \Lambda _+ \big( A_S - B_S \ppsl\big) \ ,\cr
\chi _{\rm 0A} &= \Lambda _+  \epsl \big( A_S + B_S \ppsl\big )
  \gamma _5 \ .\cr
}\eqn\eqSAAMP
$$

\chapter{Numerical Calculation}

\section{Choice of Running Coupling Constant}

Asymptotic freedom requires that at
scales $\mu^2 \gg \Lambda_{\rm QCD}^2$
the running coupling behaves
(in the one loop approximation) as
$$
\alpha(\mu^2) = {\alpha_0 \over \ln (\mu^2/\Lambda_{\rm QCD}^2)}\,,
\eqn\EQCOUPL
$$
where the constant
$\alpha_0$ depends on the number of light quark flavours ---
we work with three light flavours for which $\alpha_0 = 4\pi/9$.
Unfortunately we have no guidance on the form of the coupling outside
this deep Euclidean region.
Previous investigations of the improved ladder
approximation \refmark{\ABKMN , \JM} have simply continued the
form of \EQCOUPL\ so as to have no singularities on the real axis in
the spacelike region $\mu^2 > 0$.
However the prescriptions which were used are inadequate
for the present calculation of the Isgur-Wise function.
This is because argument $p_{\rm E}^2 + k_{\rm E}^2$ of
$\alpha(p_{\rm E}^2 + k_{\rm E}^2)$ becomes complex
when we use the BS equation \eqZEROC\ to calculate $\chi(u,x)$
at complex argument.
We thus require that the coupling be analytic and that it have
no singularities within the integration region used to perform
the analytic continuation for the boosted BS amplitude.

We adopt the following form:
$$
\eqalign{
\alpha(\mu^2) &= {\alpha_0 \over \ln (f(\mu^2/\LBS^2))}
\ ,\cr
f(x) &= x + \kappa\ \ln
  \left[\ 1+\exp\left(\,{x_0-x\over\kappa}\,\right)\ \right] \ ,\cr
}\eqn\EQanacoupl
$$
with suitable real $\kappa>0$ and $x_0>1$.
$\LBS$ is a constant analogous to $\Lambda_{\rm QCD}$ which
sets the scale of the coupling. Since we determine the value of
$\LBS$ mainly from the infra-red behaviour of the coupling,
we should not expect it to agree with $\Lambda_{\rm QCD}$ which
is usually fixed using data at much higher energies.
Notice that $f(x)\sim x$ for large
positive $x$, so that the leading asymptotic behaviour of the
coupling is correct regardless of the value of $\LBS$.

{}From eq.\EQanacoupl\ one can show that $\alpha$ tends to the constant
value $\alpha_{\rm max} = \alpha_0/\ln x_0$ in the low energy
(or timelike) region $x \ll x_0$.
We use the values
$$
x_0 = 1.01,\ 1.05\ \hbox{and}\ 1.10,
$$
corresponding to $\alpha_{\rm max} = 140.3,\ 28.6\ \hbox{and}\ 14.6$
respectively.

Next we determine $\kappa$.
$f(x)$ has branch point singularities at
$$
x_{\rm s}\equiv x_0+i\pi\kappa(2n+1)\ ,\quad n\in{\bf Z}\ ,
\eqn\EQsingularity
$$
with the branch cuts extending horizontally to the left
if the principal value of the logarithm is taken; these singularities
give rise to corresponding singularities in $\alpha(\mu^2)$.
So $\kappa$ must be sufficiently large that these singularities
and branch cuts lie outside the integration region used to calculate
the BS amplitude $\chi(u,x)$ with complex arguments $u$ and $x$.
{}From eq.\eqCOPARA, the complex momentum squared $p_{\rm E}^2$ is given by
$$
p_{\rm E}^2 = u'^2 - 2im(\,t-1\,)u' + x'^2
    - 2mx'\cos\theta \sqrt{\,t^2-1} + 2m^2(\,t-1\,) \ ,
\eqn\momentumsquared
$$
where $u'$ and $x'$ run over real values.
We can avoid the singularity in
$f\big(\,(p_{\rm E}^2+k_{\rm E}^2)/ \LBS^2\, \big)$ if,
when ${\rm Re}\,(\,p_{\rm E}^2 + k_{\rm E}^2\,)/\LBS^2 = x_0$,
$$
  |{\rm Im}\,(\,p_{\rm E}^2 + k_{\rm E}^2\,)|
  = 2m(t-1)u' < \pi \kappa\LBS^2 \ ,
\hbox{\ for all}\quad u'\,{,}\,x'\,{,}\,k_{\rm E}^2\, > \,0 \ .
\eqn\conditionK
$$
When ${\rm Re}\,(\,p_{\rm E}^2 + k_{\rm E}^2\,)/\LBS^2 = x_0$,
$$
\eqalign{
  x_0 \LBS^2 &= {\rm Re}\,(p_{\rm E}^2 + k_{\rm E}^2) \cr
	&= k_{\rm E}^2 + u'^2 + (\,x'-m\sqrt{\,t^2-1}\cos\theta\,)^2
	+ m^2 (\,2t-2\ - (t^2-1)\cos^2\theta\,) \cr
	&\geq u'^2 - m^2(t-1)^2 \ , \cr
}\ee
$$
so that
$$
u' \leq u'_{\rm max} \equiv
    \sqrt{\,x_0\LBS^2 + m^2 (\,t-1\,)^2 \,} \ .
\ee
$$
Thus we require
$$
  \kappa > {2(t-1)\over\pi}{m\over\LBS}
	\sqrt{\,x_0 + {m^2\over\LBS^2}(\,t-1\,)^2}
\eqn\thing
$$
for the range of values of $t$ we consider. As will be seen below,
at fixed $\kappa$,
 $m/\LBS$ is determined by $x_0$ and the RHS is a function
of only $x_0$ and $t$.
Using the data for $m/\LBS$ obtained below for the three values
of $x_0$, we can estimate the RHS of \thing\ and find that
$\kappa = 0.3$ allows us to calculate the Isgur-Wise function
for $t<1.55$. As will be seen below,
this region of $t$ covers the kinematically allowed
region for the semi-leptonic $B$-decay which we are interested in.
So we fix $\kappa$ equal to 0.3 henceforth.
The reader may have noticed that there are also singularities in
eq.\EQanacoupl\  when $f(x) = 0$ or $f(x)=1$;
it is easy to see that the condition \thing\ given above is sufficient
to avoid these singularities.

\section{Fixing the Mass Scale and the Light Quark Masses }

In order to treat dynamical chiral symmetry breaking consistently
in the ladder approximation we should use the dynamical quark
mass function obtained from the improved ladder approximation to
the SD equation, and we should use the same kernel in
both the SD and BS equations. It is easy to see that this
approach correctly leads to massless pions in the chiral
limit.\refmark{\ABKMN}

In fact, the inclusion of the running mass is plagued
with a number of technical difficulties related to the
consistency of the axial WT identity.\refmark{\KuMiWT}
In the present calculation, there is the further problem
that, after Wick rotation of the
loop momentum, the momenta flowing through the light and heavy fermion
propagators, $\SH(p+\zeta q)$ and $\SL(p-\eta q)$,
become complex necessitating values of the quark
mass function $\Sigma(x)$ with complex arguments
$x=-(p+\zeta q)^2,\ -(p-\eta q)^2$.
These are not insurmountable problems (in particular the quark mass
function with complex argument may be evaluated using the SD equation
in the same way as we calculated the boosted BS amplitude in Sect.3),
however they considerably complicate the formalism.
In this paper we therefore use fixed quark masses and
defer the discussion using the running mass function to a future
paper.

It turns out that the leading order calculations
need no particular value of the heavy quark mass $M$,
but we do have to fix the light quark mass $m$ as well
as the scale $\LBS$ used in the coupling.
In order to do this we adopt the following procedure.
First we solve numerically the improved ladder SD equation for the light
quark mass function $\Sigma(p_{\rm E}^2)$, using the running coupling
$\alpha(p_{\rm E}^2 + k_{\rm E}^2)$ given in eq.\EQanacoupl;
$$
\Sigma(x) = {3C_2 \over 4\pi} {\int\nolimits_0^\infty \llap {$dy$}}
		\ {\alpha(x+y) \over \max(x,y)}
		\ {y\Sigma(y) \over y+\Sigma(y)^2}\ . \eqn\EQsd
$$
There is no unique definition of the constituent quark mass
in terms of the quark mass function $\Sigma$;
we work with the following two definitions\refmark{\GP}
$$
\eqalign{
m &= \Sigma(m^2) \quad\hbox{type I}\ ,\cr
m &= \Sigma(4m^2) \quad\hbox{type I$\!$I}\ .\cr
}\eqn\EQmasses
$$
Next, to determine the energy scale $\LBS$,
we calculate the pion decay constant $\FPS$ using the obtained mass
function $\Sigma $ and the Pagels-Stokar formula\refmark{\PS}\ which reads
$$
\FPS ^2 = {N_c \over 2\pi^2} \int\nolimits_0^\infty dx
	\ { {x\Sigma(x)\,(\,\Sigma(x) - x \Sigma^\prime(x)/2\,)}
		\over (\, x + \Sigma(x)^2\,)^2 }\ .
\ee
$$
Since we know the Pagels-Stokar formula to agree
with the ladder exact result to $10\sim 20$\%, \refmark{\ABKMN,\KuMi}
imposing the value  $\FPS = 93\sqrt{2}$MeV allows us to fix $\LBS$.
As a consistency check we also calculate the expectation value of
the quark bilinear $\VEV{\bar\psi\psi}_{1\rm GeV}$,\refmark{\ABKMN}
using the formula
$$
\VEV{\bar\psi\psi}_{\rm 1GeV} =
	- \left({\alpha(\Lambda) \over
	\alpha({\rm 1GeV})}\right)^{9C_2\over 11N_c-2N_f}
	{N_c \over 4\pi^2}
	{\int\nolimits_0^{\Lambda^2} dx}
	\ {x \Sigma(x) \over x + \Sigma(x)^2}
\eqn\eqVEV
$$
with sufficiently large ultraviolet cutoff $\Lambda$, where $N_f=3$ is
the number of light flavours.
This value \eqVEV\ should be compared with the following
``experimental'' value
given by Gasser and Leutwyler\refmark{\GL}
$$
- \VEV{\bar\psi\psi}_{1\rm GeV}\  =
	(\ 225 \pm 25 \ \hbox{MeV}\ )^3\ .
$$

\TABLE\TABlqcd
The results obtained by this procedure are listed in Table.\TABlqcd.
\vskip .8cm
\begintable
$x_0$\|$\ \LBS$ \ |
 $\ m_{\rm type\ I} \ $ | $m_{\rm type\ II} \ $ |
$\ (-\VEV{\bar\psi\psi}_{1\,{\rm GeV}})^{1/3}\ $ \crthick
 1.01 \| 638 | 489 | 288 | 212 \cr
 1.05 \| 631 | 482 | 286 | 213 \cr
 1.10 \| 625 | 474 | 285 | 214
\endtable
\vskip .2cm
\caption{Table.\TABlqcd. Parameter values used in our calculation.
Units are MeV.}

\section{Bethe-Salpeter equation}

In order to solve the BS equation \eqZEROC\
numerically, we discretise the variables $u$ and $x$ so
that it reduces to a finite dimensional
eigenvalue problem, which we solve using a standard
linear algebra package.
Since logarithmic scale seems natural, we discretise the variables
$U \equiv \ln (u/\LBS)$ and
$X \equiv \ln (x/\LBS)$ at $N_{\rm BS}=30$ points evenly spaced
in the intervals
$$
U \in [\, -10.0 ,\, 2.5\, ] \ ,\quad
X \in [\, -4.5 ,\, 4.0\, ] \ .
\eqn\EQwidth
$$
The integration kernel $K_0(u,x;v,y)$,
given explicitly in the appendix,
has an integrable logarithmic singularity at $ (u,x) = (v,y) $
which  we avoide by using the four point average prescription
\refmark{\AKM}.
Calculation of the binding energy $E_0$ and the corresponding
BS amplitude $\chi_0(u,x)$ is straightforward, and we can
immediately calculate the decay constant $\FB$ by integrating the real
part of $A_0(u,x)$ via eq.\eqFB.
\TABLE\TABlattice
The number of sites $N_{\rm BS}$ and the support of $U$ and $X$ in
eq.\EQwidth\ are large enough to guarantee the discretisation
independence of the binding energy and the decay constant to within
1 \% --- see Table.\TABlattice.
As a consistency check, we also confirm that
for the fixed coupling hydrogen atom like case
our program gives results which agree well with
the formula $E_0 = m\alpha^2/2$ for small $\alpha$.

\vskip .8cm
\begintable
type | $x_0$ |
$N_{\rm BS}\ $\|$\FB\sqrt{M_B}$\ |$E_0^{(0)}$|$E_0^{(1)}$\crthick
I | 1.05 | 18 \| 3468 | 935.0 | 498.3\cr
I | 1.05 | 30 \| 3484 | 931.7 | 499.4\cr
%I | 1.05 | 32 \| 0.22002 | 1.4769 | 0.79172\cr
I | 1.05 | 34 \| 3486 | 931.4 | 499.3
\endtable
\vskip .2cm
\caption{Table.\TABlattice.  $N_{\rm BS}$ independence.
	$E_0^{(0)}$ and $E_0^{(1)}$ are binding energies of ground
	state and first excited state, respectively.
	Units are (MeV)$^{3/2}$ for $\FB\sqrt{M_B}$ and MeV for
        $E_0^{(0)}$ and $E_0^{(1)}$.}
\vskip10pt

\FIG\FIGNormFb{Figure \FIGNormFb. Integrands of normalisation
	\eqTHENORM\ and $\FB$ \eqFB.
	$N_{\rm BS}=30$ and $x_0=1.05$. The upper $9/10$ of the
	figures are clipped.}
As can be seen in figure \FIGNormFb,
the main supports of the integrands of normalisation \eqTHENORM\ and decay
constant \eqFB\ are included in the ranges given in \EQwidth.
Unfortunately we can also see that these supports
extend far into the infra-red --- much further than
in the case of the pion \refmark{\ABKMN} --- and consequently
our results depend on the infra-red behaviour of the coupling.
We are therefore forced to include $x_0$ as a parameter in our
model.
\vskip10pt
\figmark{\FIGNormFb}

\TABLE\TABdep
Our numerical results for a range of values of $x_0$
are shown in Table.\TABdep.
We prefer the
value $x_0 = 1.05$ for the simple reason that at this
value the shape of the running coupling most resembles the
form used in our previous work\foot
{Indeed, we have calculated $F_B\sqrt{M_B}$ using the previous
running coupling function\refmark{\ABKMN} and confirmed that
it produces almost the
same result as the present one for the choice $x_0=1.05$.
}.\refmark{\ABKMN,\AKM,\KuMi}
Ideally we would use the excitation energies
$E_0^{(0)}-E_0^{(1)}$ to fix the parameter $x_0$,
but at present there is no experimental data for the masses of the
excited pseudoscalar or vector $B$ or $D$ mesons.
\vskip.8cm
\begintable
type | $x_0$ | light quark mass \| $F_B\sqrt{M_B}$ |
$E_0^{(0)}$ | $E_0^{(1)}$ | $E_0^{(0)}-E_0^{(1)}$ \crthick
     I | 1.01 | 489 \| 2551 | 1795 | 1071 | 724 \cr
     I | 1.05 | 482 \| 3468 |  935 |  498 | 437 \cr
     I | 1.10 | 474 \| 4093 |  666 |  319 | 347 \crthick
I$\!$I | 1.01 | 288 \| 2052 | 1558 |  972 | 586 \cr
I$\!$I | 1.05 | 286 \| 2738 |  799 |  447 | 352 \cr
I$\!$I | 1.10 | 285 \| 3205 |  566 |  279 | 287
\endtable
\vskip .2cm
\caption{Table.\TABdep. $x_0$ dependence; $N_{\rm BS}=18$.
	Units are MeV except for $F_B\sqrt{M_B}$
	which is in (MeV)$^{3/2}$.}

\TABLE\TABFBdata
The $B$ and $D$ meson decay constants and mass difference using a finer
discretisation ($N_{\rm BS}=30$) are given in Table.\TABFBdata\ where
these meson masses are taken from Ref.[\PDG].
\vskip .8cm
\begintable
type | $x_0$ |mass \| $F_{B(5279)}$ |
  $F_{D(1869)}$ | $B(\hbox{1st})-B(5279)$ \crthick
 I | 1.05 | 482 \| 48.0 | 80.6 | 432 \cr
 I$\!$I | 1.05 | 286 \| 37.8 | 63.4 | 350
\endtable
\vskip .2cm
\caption{Table.\TABFBdata. Results for the meson decay constants and
	mass difference with $N_{\rm BS} = 30$ and $x_0 = 1.05$.
	$B(\hbox{1st})$ denotes 1st excited state of
        pseudoscalar $B$ meson. Units are MeV.}
\noindent
As can be seen, our result for $\FB$ is smaller than $F_\pi$,
and is much smaller than other values obtained using QCD sum rules,
potential models and lattice simulations (see the summary in the paper
by Rosner\refmark{\Rosner}) --- we note however that one gluon
exchange models tend to give small values,\refmark{\Suzuki}\
although even these are somewhat larger than our results.

\section{The Isgur-Wise function : $\xi(t)$}

After evaluating the BS amplitude of the ground state, we can now apply
the formalism of section 4 and calculate
the Isgur-Wise function $\xi(t)$ by numerically evaluating the integral
eq.\eqISGWISINT.
This integral itself is three dimensional, over $u',x'$ and $\cos
\theta$, but as explained in section 4, the calculation of the initial
BS amplitudes $A(u,x)$, $B(u,x)$ with complex arguments $u$ and $x$
requires the evaluation of the BS equation \eqZEROC\ which involves a
further two dimensional integral over $v$ and $y$.
So in fact we are calculating a five dimensional integral, which takes
a lot of computer time when the number $N_{\rm BS}$ of points on which
the arguments $u$, $x$, $v$ and $y$ of the BS amplitudes are
discretised becomes as large as 30.
In order to save computer time, we evaluate the angle integral over
$\cos\theta$ using the Gauss-Legendre formula for numerical
integration which is known to give rather precise values with using a
relatively small number of data points $\NGL$.

The main experimental interest at the moment is in semi-leptonic decay
processes such as $B \rightarrow D^{(*)}l\nu$,
since these should allow calculation of the Kobayashi-Maskawa matrix
element $V_{\rm cb}\refmark{\KM}$.
For these processes, the invariant mass between $l$ and $\nu$ must be
positive, and this bounds
$t$ from above
$$
t \leq { M_B^2 + M_{D^{(*)}}^2 \over 2M_BM_{D^{(*)}} }
       \simeq 1.5 - 1.6\ .
\ee
$$
\FIG\FIGXI{Figure \FIGXI. Isgur-Wise functions calculated with $N_{\rm
	BS}=30$, $N_{\rm GL}=10$ and $x_0=1.05$ for both type I and
	type I$\!$I masses.
}
We have chosen the parameter $\kappa$ in our running coupling so that we
may evaluate the Isgur-Wise function within this region.
The result can be seen in figure \FIGXI, where we have shown the
Isgur-Wise function for the two cases of type I and type II masses.
Figure \FIGXI\ is the main numerical result in  this paper.
\figmark{\FIGXI}
\FIG\FIGXIdepI{Figure \FIGXIdepI.
	$x_0$ dependence of the Isgur-Wise function
	with type I mass, $N_{\rm BS}=18$ and $N_{\rm GL}=10$.}
\FIG\FIGXIdepII{Figure \FIGXIdepII.
	$x_0$ dependence of the Isgur-Wise function with type I$\!$I
	mass,$N_{\rm BS}=18$ and $N_{\rm GL}=10$.}

In figures \FIGXIdepI\ and \FIGXIdepII, we show how the result depends
on the choice of the parameter $x_0$ in the running coupling constant,
for both the type I and type II masses.
These data are based on
the BS solutions on a coarser lattice with $N_{\rm BS}=18$ points
and $\cos\theta$
integration evaluated by $N_{\rm GL}=10$ Gauss-Legendre formula.
We see that the result is almost independent of $x_0$ for the type I
case, whereas there is some dependence for the type II case.
\figmark{\FIGXIdepI}
\figmark{\FIGXIdepII}

Here a comment may be in order on some \lq fluctuations\rq\ of our data
points.
We observe \lq large\rq\ fluctuations occur for instance, at $t=1.5
(x_0=1.01)$ point in Fig.\FIGXIdepI, $t=1.18 (\hbox{type I})$ point in
Fig.\FIGXI\ and so on.
We do not yet understand the precise origin of this phenomenon.
But we suspect that some \lq coherence\rq\ is occurring between the
Gauss-Legendre formula and the discretisation for the BS amplitude
data.
Indeed, if we increase $N_{\rm GL}$ of the Gauss-Legendre formula,
the Isgur-Wise function calculated by using the same BS data becomes
more smooth as a whole, {\it but} the places at which such \lq
large\rq\ fluctuations occur change.
Actually, for example, the large deviation at $t=1.5(x_0=1.01)$ in
Fig.\FIGXIdepI\ does not appear when $N_{\rm GL}=6$.
(Therefore the seeming \lq oscillation\rq\ observed in Fig.\FIGXIdepII\
around $t=1.3\sim 1.5$ will also be a fake.)
If we are allowed to average all the data obtained with various
$N_{\rm GL} (\geq 6)$, we obtain very smooth curves which coincide
with the cited $N_{\rm GL}=10$ data with fluctuations smoothed out.

The notable feature of these figures \FIGXIdepI\ and \FIGXIdepII\ is
that the charge radius of the Isgur-Wise function
$$
\rho^2 \equiv -{d\xi(t) \over dt}{\Big\vert}_{t=1}
\ee
$$
is as large as 2.0 for type I mass or 1.8 for type II, and is
quite insensitive to the choice of $x_0$.
These values of $\rho^2$ are certainly larger than the
usual predictions of
$\rho^2 = 1.1\sim1.3$.\refmark{\Rosner, \JHD, \MRR}
In our model
the slope of the Isgur-Wise function does not decrease to 1.1 until
region $t=1.1 \sim 1.3$.
If this feature of our result is correct,
it would imply that
$V_{\rm cb}$ as extracted from the experimental
data\refmark{\PB, \KS, \BSW, \GISW, \HMW, \CKP}\
has been underestimated by
several percent.

To see this more explicitly, we fit our Isgur-Wise function
$\xi(t)$ of figure \FIGXI\ to the ARGUS data\refmark{\ARGUS} ( with
setting their parameter $\tau_B = 1.32\ \hbox{ps}$ )
by adjusting $V_{\rm cb}$
so as to minimise
$\chi^2 = \sum_n \left(\,|V_{\rm cb}|\,\xi(t_n) - f_n \right)^2
/ \sigma_n^2$,
where $f_n$ and $\sigma_n$ are the experimental data and standard
deviations at $t=t_n$.
\FIG\FIGKMxi{Figure \FIGKMxi. $\xi(y) \cdot |V_{\rm cb}|$ versus $y$.
	$x_0 = 1.05$ and both type I and II masses are used.
}
The result is shown in figure \FIGKMxi.
We find
$$
|V_{\rm cb}| =
\cases{
	0.0503  & for type I case \cr
	0.0437  & for type II case \cr }\ ,
\ee
$$
and see that
the type I case gives a better overall fit to the ARGUS data.
\figmark{\FIGKMxi}

Finally, we use our Isgur-Wise function with type I mass
to estimate the $\rho^2$ parameters in various functional forms of
$\xi(t)$ proposed so far.\refmark{\PB, \KS, \GISW, \NR, \Neu, \Neub}
\FIG\FIGXICOMP{Figure \FIGXICOMP. Least squares fits of the
	four theoretical models $A$, $B$, $C$ and $D$ to our numerical
	data.}
\TABLE\TABxiparam
The functional forms and the corresponding charge radii $\rho^2$
which give  the best
fit to our data are given in Table.\TABxiparam, and the resulting
curves are shown in figure \FIGXICOMP.
The model C gives the least $\chi^2$, although the models B and C also
give good fits.
\vskip .8cm
\begintable
model \| function form | $\rho^2$ \crthick
A \| $1-\rho^2(t-1)$ | 1.31 \cr
B \| $\frac{2}{t+1}\exp\left[ -(2\rho^2-1)\frac{t-1}{t+1}\right]$
					| 2.16 \cr
C \| $\left(\frac{2}{t+1}\right)^{2\rho^2}$ | 2.03 \cr
D \| $\exp[-\rho^2(t-1)]$ | 1.87
\endtable
\vskip .2cm
\caption{Table.\TABxiparam. Various parametrisations of the Isgur-Wise
	function.
	The charge radii are extracted from our $\xi(t)$
	with $N_{\rm BS}=30$, $x_0=1.05$ and type I mass in
	Fig.\FIGXI.
}
\figmark{\FIGXICOMP}

\ACK

The authors would like to thank Machiko Hatsuda for collaboration
in the early stages of this work.
We also thank Masayasu Harada for help with the numerical work.
T.K.~is supported in part by the Grant-in-Aid for Scientific Research
(\#04640292) from the Ministry of Education, Science and Culture.
M.G.M.~thanks the European Community for support.

%---------------------------------------------------------------------

\APPENDIX{A}{\ :\ \ BS Kernel After Angle Integration}

After performing three-dimensional angle integration over
$\cos\theta = {\bf k}\cdot {\bf p}/\abs{\bf k}\abs{\bf p}$,
the zeroth order kernel $K_0(p,k)$ is found to be given by
$$
K_0(p,k) =
\mymatrix{I_1-(u-v)^2I_2}
{i(u-v)(y^2I_2-\Itpk)}
{-i(u-v)(x^2I_2-\Itpk)}
{\Iopk - (u-v)^2\Itpk} \ ,
\eqn\eqZEROKERNEL
$$
where
$$
\eqalign{
&I_1\equiv \int _{-1}^1 d\cos\theta {1\over -(k-p)^2}
  = {1\over 2xy}\ln
  \left({(x+y)^2+(u-v)^2\over (x-y)^2+(u-v)^2}\right) \cr
&I_2\equiv \int _{-1}^1 d\cos\theta {1\over (k-p)^4}
  = {2\over \big((x+y)^2+(u-v)^2\big)\big((x-y)^2+(u-v)^2\big)} \cr
&\Iopk\equiv \int _{-1}^1 d\cos\theta
{{\bf k}\cdot{\bf p}\over -(k-p)^2}
= {x^2+y^2+(u-v)^2\over 2} I_1 - 1 \cr
&\Itpk\equiv \int _{-1}^1 d\cos\theta
{{\bf k}\cdot{\bf p}\over (k-p)^4}
= {1\over 2}\big( (x^2+y^2+(u-v)^2)I_2 - I_1 \big) \ .\cr
}\ee
$$

\refout
\figout

\draw{Fig.\FIGNormFb}{\epsfbox{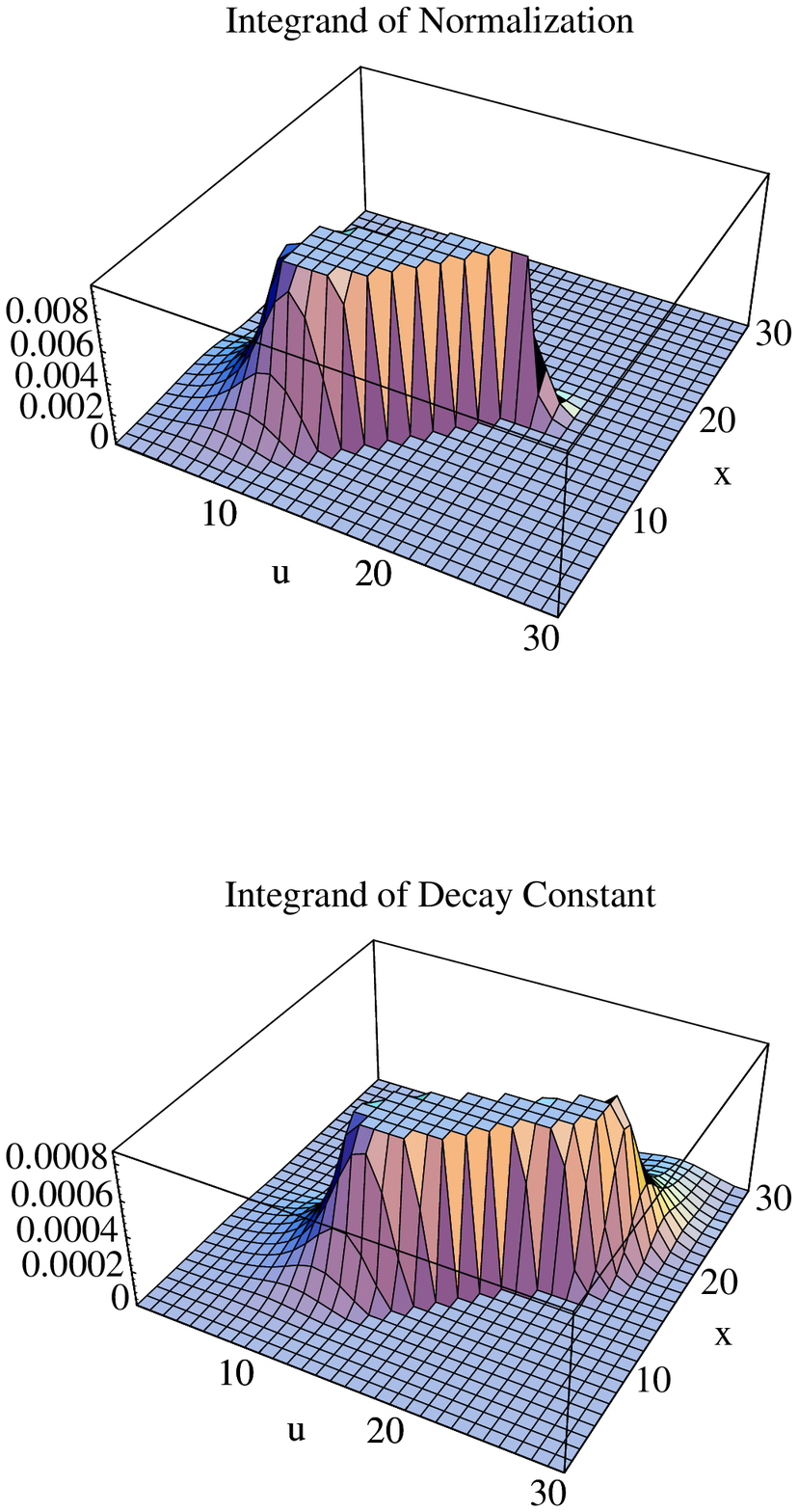}}
\draw{Fig.\FIGXI}{\epsfbox{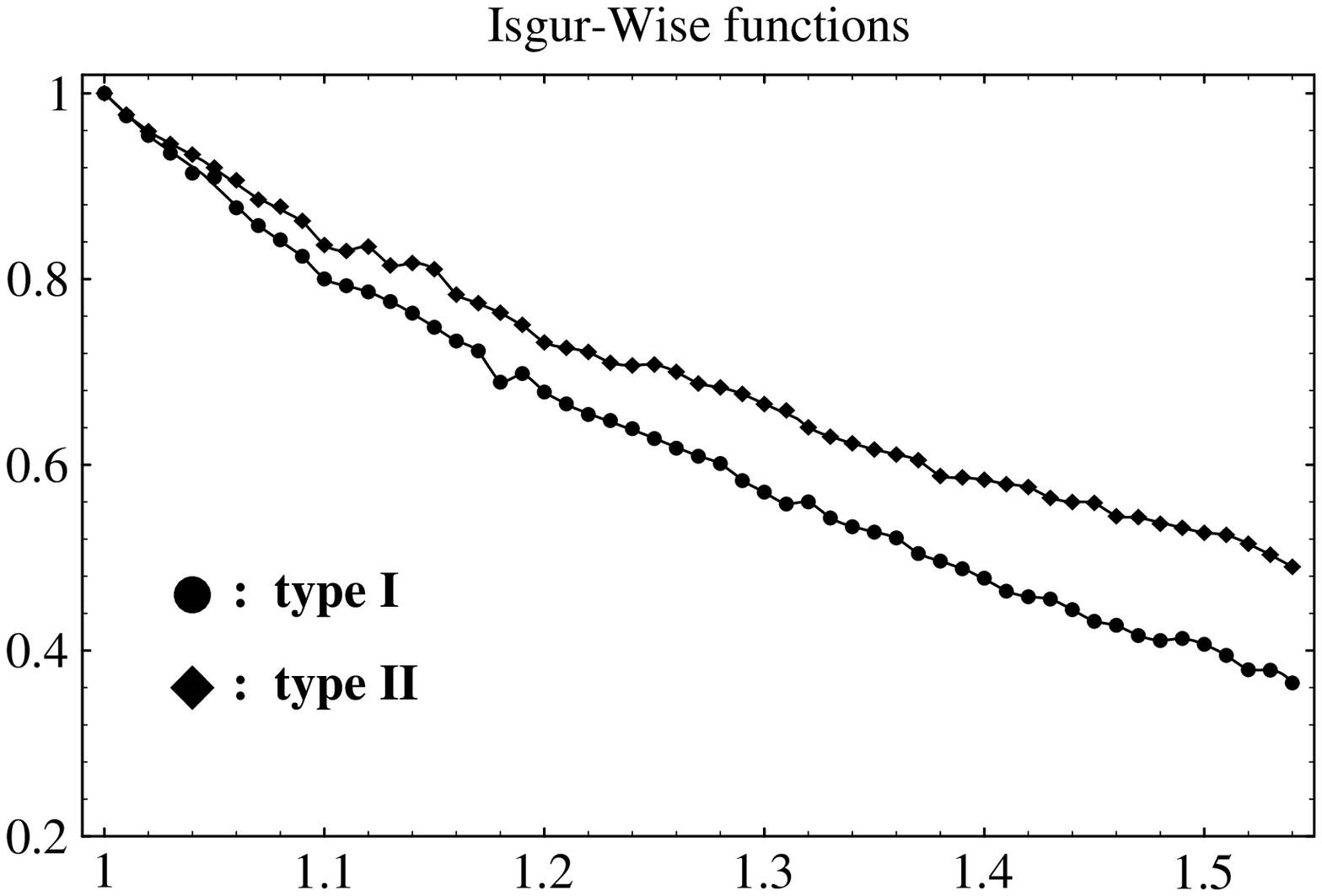}}
\draw{Fig.\FIGXIdepI}{\epsfbox{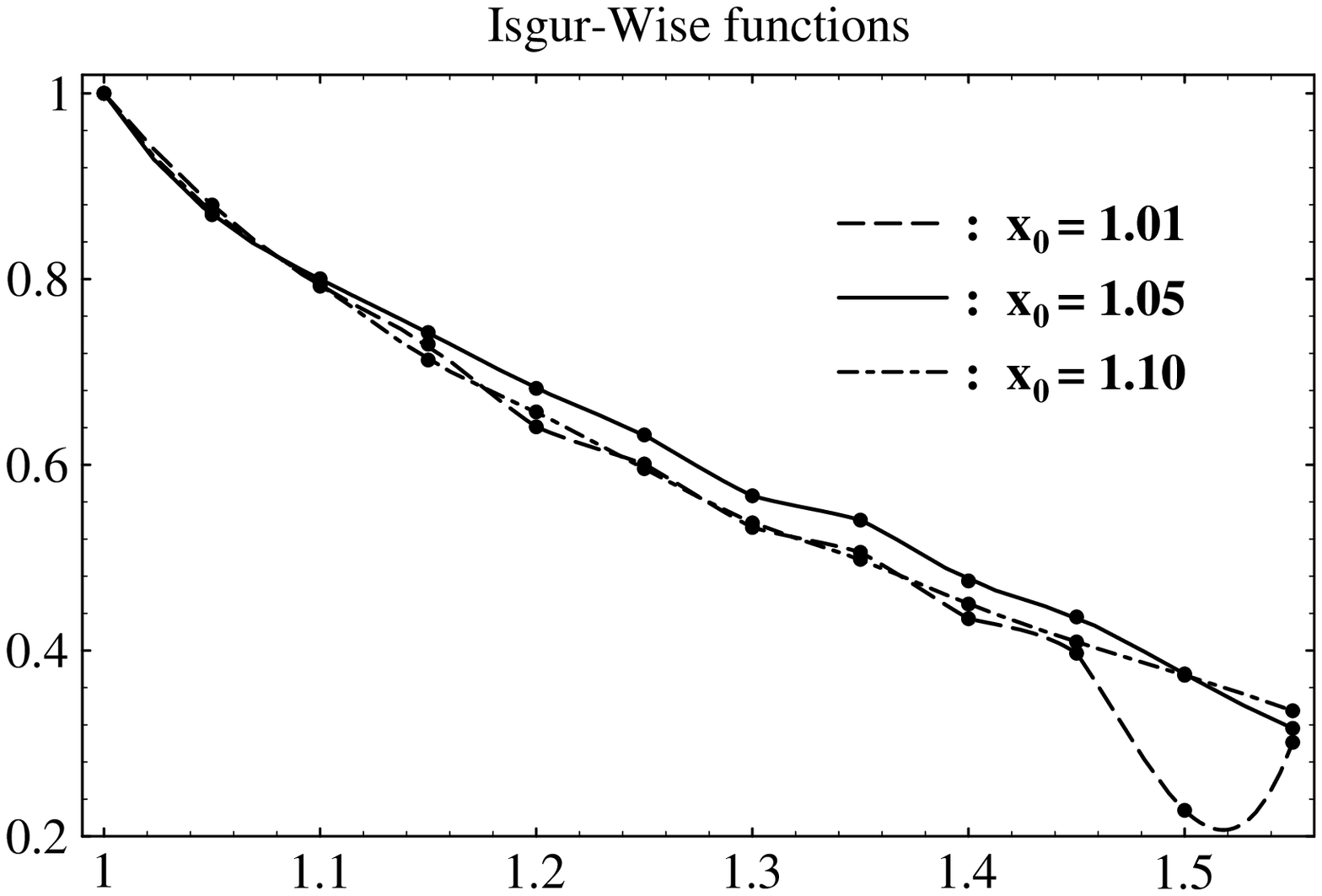}}
\draw{Fig.\FIGXIdepII}{\epsfbox{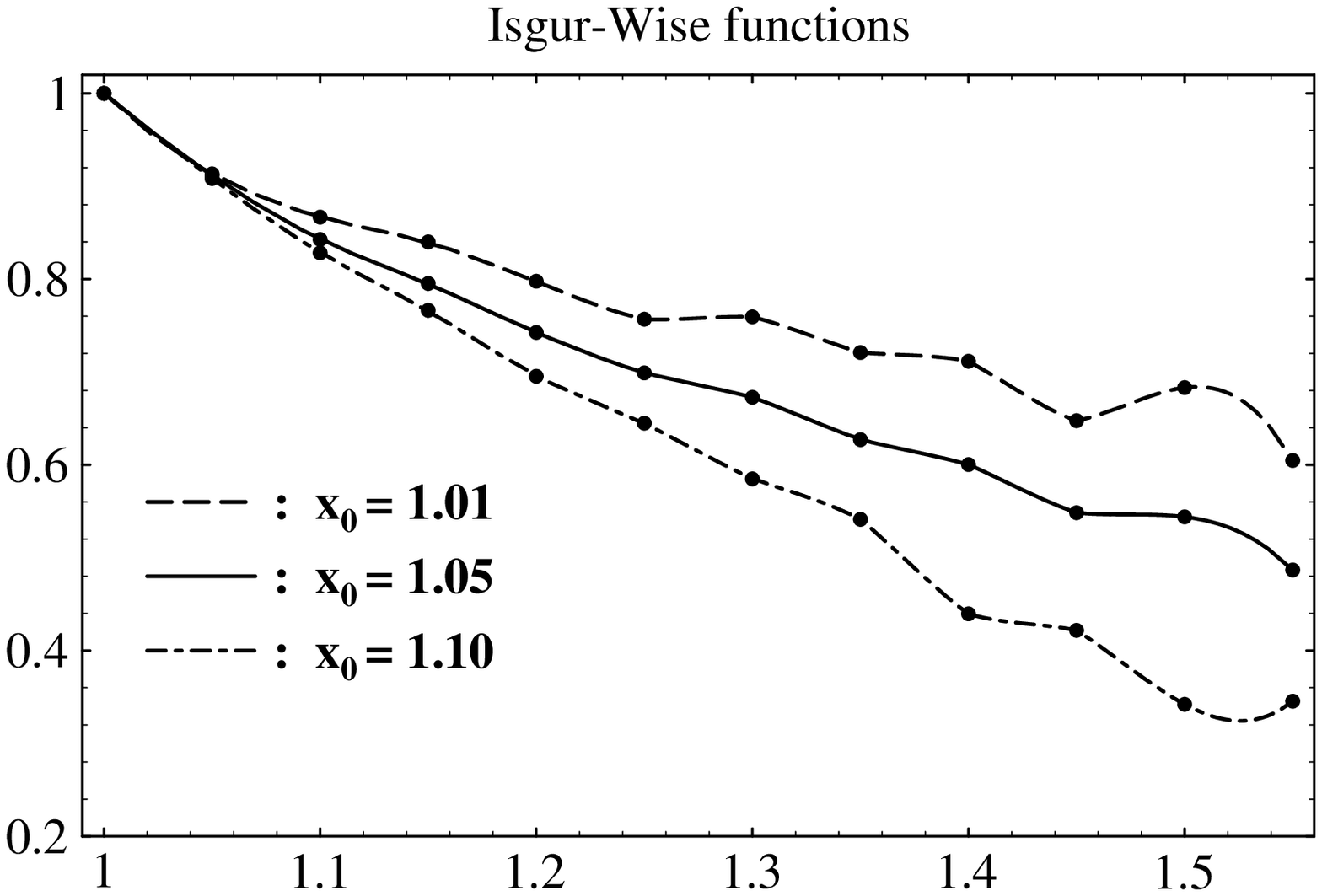}}
\draw{Fig.\FIGKMxi}{\epsfbox{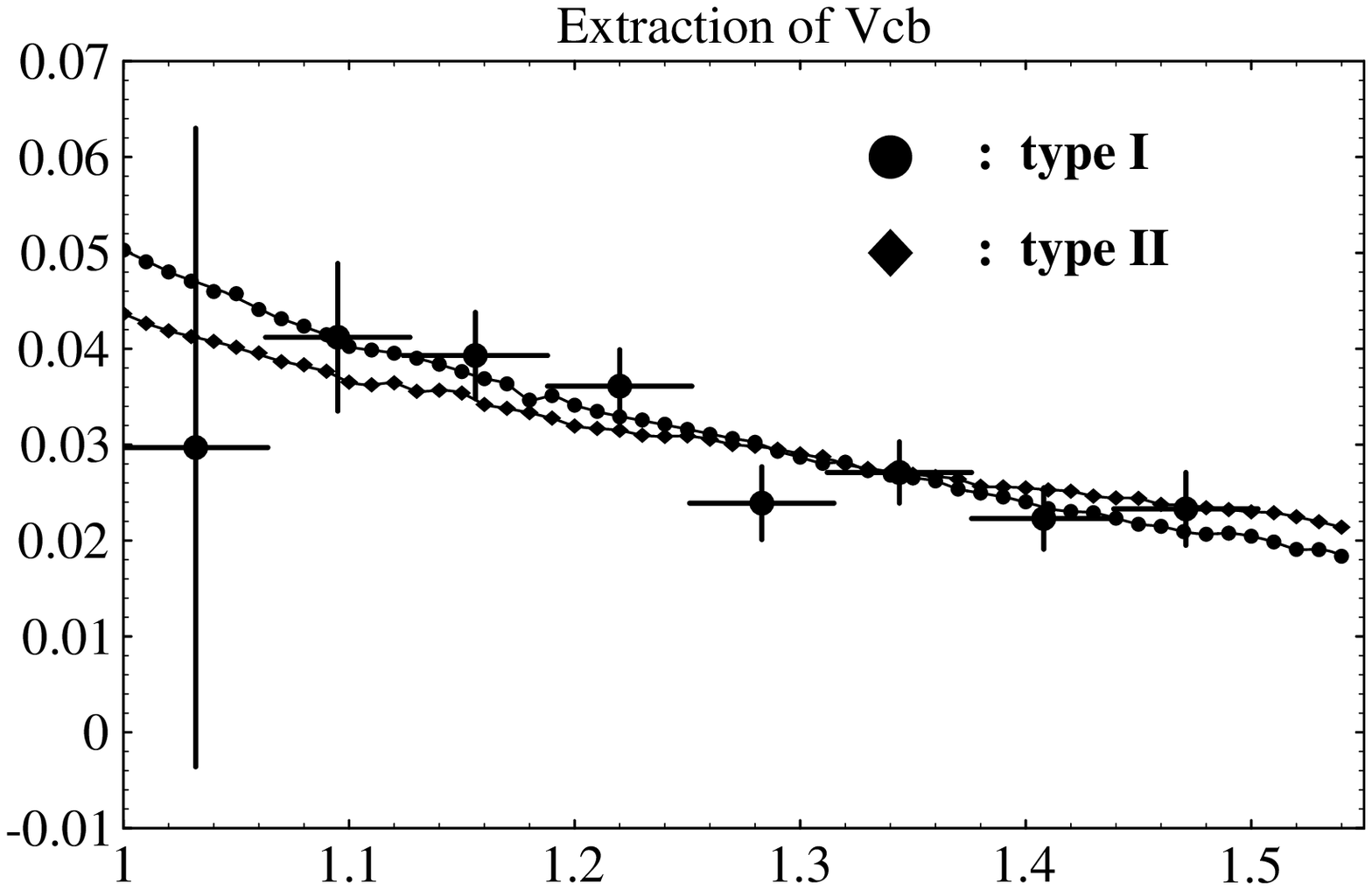}}
\draw{Fig.\FIGXICOMP}{\epsfbox{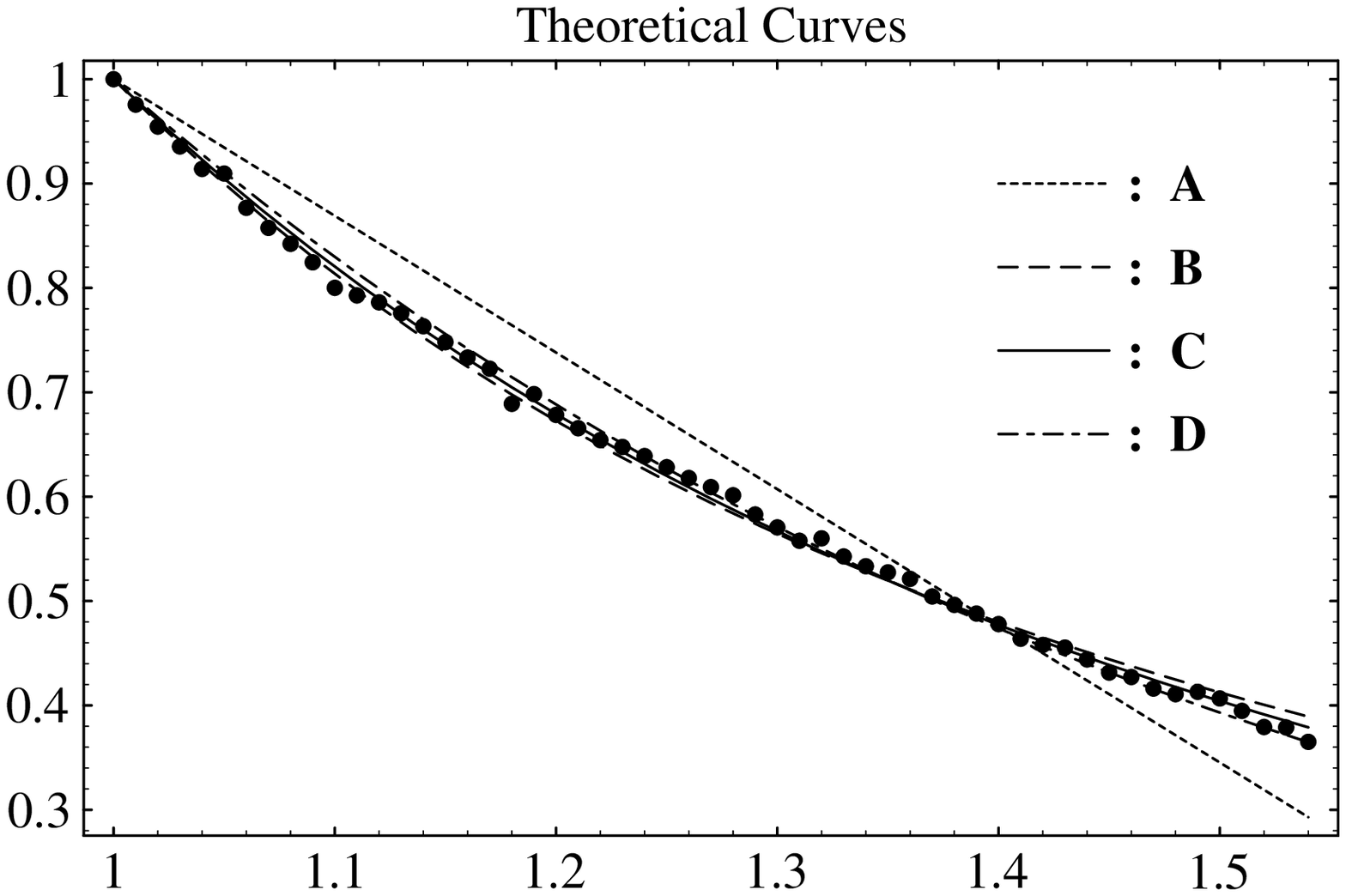}}

\bye\bye